# Continuous Mental Effort Evaluation during 3D Object Manipulation Tasks based on Brain and Physiological Signals

Dennis Wobrock[1,2], Jérémy Frey[2,3,4], Delphine Graeff[1], Jean-Baptiste de la Rivière[1], Julien Castet[1], Fabien Lotte[2,3]

1 Immersion SAS, Bordeaux, France
2 Inria Bordeaux Sud-Ouest, Talence, France
3 LaBRI, Talence, France
4 Université de Bordeaux, Bordeaux, France
dennis.wobrock@ensc.fr,
{jeremy.frey,fabien.lotte}@inria.fr,{delphine.graeff,
jb.delariviere,julien.castet}@immersion.fr

**Abstract.** Designing 3D User Interfaces (UI) requires adequate evaluation tools to ensure good usability and user experience. While many evaluation tools are already available and widely used, existing approaches generally cannot provide continuous and objective measures of usability qualities during interaction without interrupting the user. In this paper, we propose to use brain (with ElectroEncephaloGraphy) and physiological (ElectroCardioGraphy, Galvanic Skin Response) signals to continuously assess the mental effort made by the user to perform 3D object manipulation tasks. We first show how this mental effort (a.k.a., mental workload) can be estimated from such signals, and then measure it on 8 participants during an actual 3D object manipulation task with an input device known as the CubTile. Our results suggest that monitoring workload enables us to continuously assess the 3DUI and/or interaction technique ease-of-use. Overall, this suggests that this new measure could become a useful addition to the repertoire of available evaluation tools, enabling a finer grain assessment of the ergonomic qualities of a given 3D user interface.

## 1  Introduction

3D User Interfaces (UI) and systems are increasingly used in a number of applications including industrial design, education, art or entertainment [3,11]. As such, 3DUI and interaction techniques can be used by many different users with many varying skills and profiles. Therefore, designing them requires adequate evaluation tools to ensure good usability and user experience for most targeted users [2,11]. To do so, a number of evaluation methods has been developed including behavioral studies, testbeds, questionnaires and inquiries, among others [8,3,11]. This resulted in the design of more relevant, efficient and easy-to-use 3DUI.

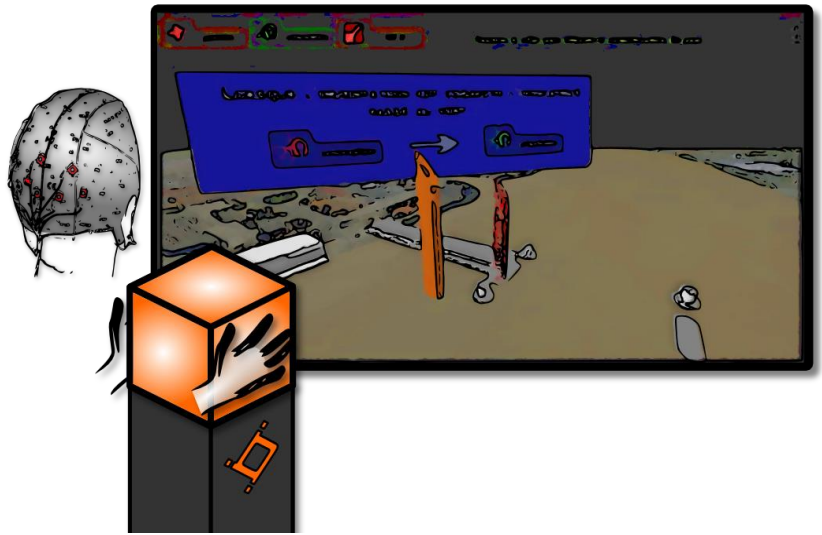

**Fig. 1.** Schematic view of a user performing 3D manipulations tasks with the CubTile input device. His/her mental effort are monitored based on brain signals (ElectroEncephaloGraphy).

Nevertheless, there is still a lot of room for improvements in the currently used evaluation methods. In particular, traditional evaluation methods could either be ambiguous, lack real-time recordings, or disrupt the interaction [8]. For instance, although behavioral studies are able to account in real-time for users' interactions, they can be hard to interpret since measures may not be specific, e.g., a high reaction time can be caused either by a low concentration level or a high mental workload [10]. Questionnaires and other inquiry-based methods such as "think aloud" and focus group all suffer from the same limitation: resulting measures are prone to be contaminated by ambiguities [19], social pressure [23] or participants' memory limitations [13].

For instance, a useful UI evaluation measure is the user's mental workload, i.e., the pressure on the user's working memory, which is typically measured using the NASA-TLX post-hoc questionnaire [10]. Even though it can be used to assess users' preferences regarding UI [12], NASA-TLX being a post-experiment measure, this is only a subjective and global measure that cannot inform on where and when the user experienced higher or lower workload. There is therefore a need for more objective (or more precisely "exocentric", as defined in [8]) and continuous measures of the usability qualities of 3DUI that do not interrupt the user during interaction.

In order to obtain such measures of the user's inner-state during interaction, a recent promising research direction is to measure such states based on brain signals – e.g., from ElectroEncephaloGraphy (EEG) – and physiological signals – e.g., from heart rate measurements or skin's moisture – acquired from the user during interaction [8]. Indeed, there are increasing evidence that the mental states that can be relevant for 3DUI evaluation, like mental workload [17], can be estimated from brain and physiological signals [8,20]. Interestingly enough, some recent works have started to use brain signal based measures of workload to compare 2D visual information displays [22,1]. However, to the best of our knowledge, estimating mental workload from both brain and physiological signals has never been explored to evaluate 3DUI, although it could provide relevant evaluation metrics to complement the already used ones. Indeed, previous works were focused on evaluating workload levels based on brain signals during 2D visualizations, thus with more passive users [22,1]. 3D interaction tasks are more complex for the user since 1) the user is actively interacting with the application, and not as passively observing it, and as such should decide what to do and how to do so, and 2) perceiving and interacting with a 3D environment is more cognitively demanding than perceiving and interacting with a 2D one, since it required the user to perform 3D mental rotation tasks to successfully manipulate 3D objects or to orientate him/herself in the 3D environment. Therefore, as compared to existing works which only explored passive 2D visualizations, monitoring mental workload seems more relevant during 3D manipulation tasks, since the user is more likely to experience pressure on his/her cognitive resources. Therefore, evaluating the resulting changes in workload levels seems even more necessary to ensure the design of usable 3DUI. Moreover, the active role of the user during 3D interaction tasks (as compared to more passive visualizations) and the higher cognitive demand as well as the richer visual feedback resulting from the use of a 3D environment means that EEG and physiological signals will be substantially different and more variable as compared to those measured during 2D visualization tasks. Finding out whether they can still be used to estimate workload levels in this context is therefore a challenging and relevant question to explore.

Therefore, in this paper, we propose to assess the mental effort (i.e., the mental workload) made by the user during 3D object manipulation tasks, based on brain (EEG) and other physiological signals. We notably propose a method to estimate workload levels from both EEG, ElectroCardioGram (ECG) and Galvanic Skin Response (GSR) signals, and we study mental workload levels during a 3D docking task in a pilot study (see Figure 1). Our results show that this approach can provide useful information about how users learn to use the 3DUI and how easy-to-use it is.

## 2 Methods

To continuously monitor workload levels during 3D interaction, we first propose an approach to estimate such workload levels from EEG, ECG and GSR signals, described hereafter. We then present how to use the resulting workload estimator to evaluate 3D object manipulation tasks and the corresponding study we conducted.

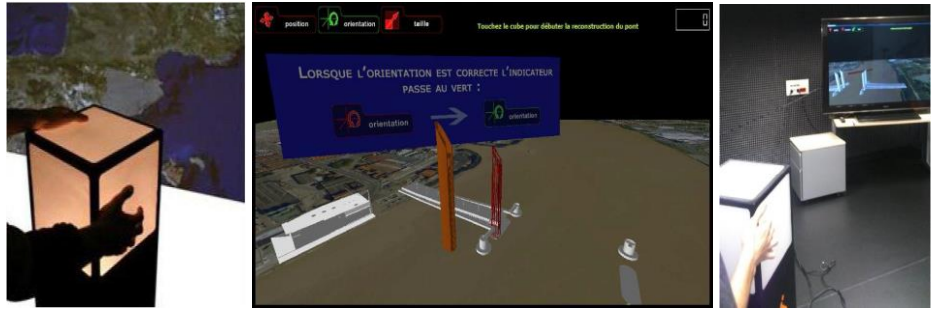

**Fig. 2.** *Left:* The CubTile interaction device. *Center:* The bridge building application. *Right:* The bridge building application controlled with the CubTile and used to analyze workload levels during 3D docking tasks.

**Measuring and Estimating Workload Levels**

Estimating workload levels from EEG, ECG and GSR signals first requires a ground truth data set with such signals labeled with the corresponding user's mental workload, in order to calibrate and validate our workload estimator. Based on this ground truth signal data set, we propose signal processing and machine learning tools to identify the user's workload level. They are described hereafter.

**Inducing mental workload to calibrate the estimator.**

To obtain a ground truth signal data set to calibrate and validate our workload estimator, we induced 2 different workload levels in our participants. To do so, we had them perform cognitive tasks, the cognitive difficulty of which being manipulated using a protocol known as the N-back task, a well-known task to induce workload (see, e.g., [17]). With such a task, users saw a sequence of letters on screen, the letters being displayed one by one, every 2 seconds. For each letter the user had to indicate whether the displayed letter was the same one as the letter displayed N letters before or was different, using a left or right mouse click respectively. Each user alternated between "easy" blocks with the 0-back task (the user had to identify whether the current letter was the letter 'X') and "difficult" blocks with the 2-back task (the user had to identify whether the current letter was the same letter as the one displayed 2 letters before). Each block contained 60 letters presentations and each participant completed 6 blocks, 3 blocks for each workload level (0-back vs 2-back). Therefore, 360 calibration trials (i.e., one trial being one letter presentation) were collected for each user, with 180 trials for each workload level ("low" versus "high").

**Measuring brain and physiological signals.**

During all experiments, EEG signals were acquired using 30 electrodes located all over the scalp (positions C6, CP4, CPz, CP3, P5, P3, P1, Pz, P2, P4, P6, PO7, PO8, Oz, F3, Fz, F4, FT8, FC6, FC4, FCz, FC3, FC5, FT7, C5, C3, C1, Cz, C2, C4), using a 32-channels g.USBAmp (g.tec, Austria). ECG and GSR were measured using a

BITalino acquisition card [28]. 3 ECG sensors were placed on the user's torso, and 2 GSR sensors on the user's index and middle fingers from the active hand. All EEG and physiological sensors were acquired using the OpenViBE platform [25].

**Signal processing tools to detect mental effort.**

In order to estimate mental workload from brain and physiological sensors, we used a machine learning approach: the measured signals were first represented as a set of descriptive features. These features were then given as input to a machine learning classifier whose objective was to learn whether these features represented a low workload level (induced by the 0-back task) or a high workload level (induced by the 2-back task). Once calibrated, this classifier can be used to estimate workload levels on new data, which we will use to estimate mental effort during 3D object manipulation tasks. From the signals collected during the N-back tasks described above, we extracted features from each 2-seconds long time window of EEG and physiological signals immediately following a letter presentation, as in [17]. We used each of these 2-seconds long time windows as an example to calibrate our classifier. Note that a classifier was calibrated separately for each participant, based on the examples of brain and physiological signals collected from that participant. Indeed, EEG signals are known to be very variable between participants, hence the need for user-specific classifiers to ensure maximal EEG classification performances [5,17].

We used the EEGLab software [4] to process EEG signals. We filtered the signals in the Delta ($1-3$ Hz), Theta ($4-6$ Hz), Alpha ($7-13$ Hz), Beta ($14-25$ Hz) and Gamma ($26-40$ Hz) bands, as in [17]. For each band, we optimized a set of 6 Common Spatial Patterns (CSP) spatial filters (i.e., linear combinations of the original EEG channels that lead to maximally different features between the two workload levels) [24,15]. For each frequency band and spatial filter, we then used as feature the average band power of the filtered EEG signals. This resulted in 30 EEG features (5 bands × 6 spatial filters per band). Note that high frequency EEG is likely to be contaminated by muscle activity (ElectroMyoGraphy - EMG) from the user's face or neck [6,9]. As such, we explored EEG-based workload estimation based on low frequencies only (Theta, Delta, Alpha) and both low and high frequencies (Theta, Delta, Alpha, Beta, Gamma).

This signal processing approach is the one we used to discriminate workload levels from EEG signals between 0-back and 2-back tasks, i.e., within the same context on which the workload estimator was calibrated. However, it is known that EEG signals change between different contexts, due, e.g., to the different user's attention and involvement that the context triggers or to different sensory stimulations (e.g., different visual inputs) that change brain responses and thus EEG signals. This means that a workload estimator calibrated in a given context will have poorer performances (i.e., will estimate an erroneous workload level more often) when applied to a different context [17]. In our experiment, the final application context, i.e., 3D objects manipulation, is very different from the calibration context, i.e., the N-back tasks. Indeed, during the N-back tasks the user is moving very little as he/she is only performing mouse clicks, and exposed to very little visual stimulations as the N-back task only involves the display of white letters on a black background. On the contrary, manipu-lating 3D objects means that the user will be moving more and would be exposed to very rich visual stimulations. As such, a workload estimator simply calibrated on the N-back tasks and applied to the 3D object manipulation tasks is very likely to give very poor results or even to fail. Therefore, we modified the above mentioned signal processing approach to make it robust to EEG signal changes between the two contexts. In particular, rather than using basic CSP spatial filters, we used regularized CSP spatial filters [14] that are robust to changes between calibration and use contexts. To do so, based on [27], we estimated the EEG signal covariance matrix from the calibration context (N-back tasks) and from the use context[1] (3D object manipulation tasks), and computed the Principal Components (PC) of the difference between these two matrices. These PC represent the directions along which EEG signals change between calibration and use. These directions are then used to regularize the CSP spatial filters as in [27], to ensure that the obtained spatial filters are invariant to these EEG signals changes.

From ECG signals we extracted the Heart Rate (HR) and 2 features from the Heart Rate Variability (HRV), namely the low frequency HRV ($< 0.1$ Hz) and the Root Mean Square of Successive Differences, as in [16], using the Biosig Matlab toolbox [29]. From GSR signals we also extracted 3 features: the mean GSR amplitude, skin conductance responses (SCR, here band power between 0.5 Hz and 2 Hz) and the skin conductance level (SCL, $0.1 - 0.5$ Hz) [7].

We then used a shrinkage Linear Discriminant Analysis (sLDA) classifier [15] to learn which feature values correspond to a high or low workload level.

Note that since both ECG and GSR analyses rely on low frequencies, we had to extend the time windows from 2s to 10s when we studied those physiological signals (for instance, for HRV at 0.1Hz, 10s are needed to observe a single cycle). As such the number of trials per condition (0-back vs 2-back) in these particular scenarios were reduced from 180 down to 36.

**Estimating mental effort during 3D manipulation**

Once we have a classifier that can estimate workload levels from brain and physiological signals, we can use it to study mental effort during 3D objects manipulation tasks. With this objective in mind, we designed an experiment in which participants, equipped with the sensors described previously (EEG, ECG and GSR), had to manipulate 3D objects using an interaction device known as the CubTile [26]. In particular, participants had to perform 3D docking tasks in order to build a bridge in 3D by assembling together its different parts (see Figure 2). The following sections describe the participant population used for this experiment, the CubTile input device and the protocol of the bridge building application.

---

[1] Note that this is only possible here because we perform an offline evaluation, after the 3D manipulation tasks have been performed and the corresponding EEG signals collected. It would not be possible to use the exact same algorithm for a real-time estimation of workload during 3D objects manipulation tasks as the covariance matrix of EEG signals during these tasks is not yet fully known.

**Population and apparatus.**

8 participants (2 females, age from 16 to 29) took part in this study. They were all first-time users of the bridge building application and the CubTile (except for one participant who has used the CubTile before for another application). These participants completed 3D manipulation tasks using the CubTile. The CubTile is a multitouch cubic interface consisting in a medium-sized cube where 5 out of 6 sides are multitouch. It can sense several fingers, offers interaction redundancy and lets a user handle 3D manipulation thanks to single handed and bimanual input [26]. In particular, with the CubTile, translations can be performed by moving symmetrically two fingers each on opposite sides of the cube. Scaling can be performed by connecting or disconnecting two fingers. Finally, rotations can be performed either by rotating symmetrically several fingers each set on opposite sides of the cube or by translating asymmetrically two fingers each on opposite sides of the cube.

**Protocol.**

The experiment took place in a dedicated experimental room, in a quiet environment. When the participant entered the room, he/she was told about the experiment and then equipped with the different sensors. Then he/she participated into 6 blocks of the N-back task (except S6 who only completed 3 blocks due to technical issues), on a standard computer screen, in order to obtain calibration data to setup the workload classifier. This took approximately 15 minutes. The participant also participated in two other calibration sessions (about 15 minutes each), to calibrate two other mental states that were not used nor analyzed for this study. Once the calibration sessions were completed, the participant was asked to sit in front of the CubTile which was itself in front of a 65 inches Panasonic TX-P65VT20E screen.

The participant then had to construct the 3D bridge by assembling the bridge parts (e.g., the 4 supporting pillars and the road) one by one. In particular, the participant had to perform docking tasks, by translating, rotating and scaling the bridge parts, in order to put them at the correct location. The correct location was indicated to the user with proper 3D feedback, integrated to the 3D scene, in the form of text and color indicating how close he/she was from the correct position, scale and orientation. All the translations, rotations and scaling were controlled by the CubTile. The participant had to perform a set of 7 docking tasks:

1. Positioning the $1^{st}$ pillar, by controlling rotation, translation and scaling – repeated 3 times for different angles, sizes and positions
2. Positioning the $2^{nd}$ pillar, by controlling 2 translations, 1 rotation and scaling, while the pillar was being continuously and automatically translated along the vertical axis – repeated 4 times for different angles, sizes and positions
3. Positioning the lower half of the $3^{rd}$ pillar by controlling a crane carrying the pillar part, along 1 rotation and 1 translation (up/down) – repeated 3 times for different angles and heights
4. Positioning the upper half of the $3^{rd}$ pillar by controlling a crane along 1 rotation and 1 translation, seen from a different angle than the previous task – repeated 3 times for different angles and heights

5. Positioning the 4$^{th}$ pillar by controlling 2 translations and 1 rotation. Without warning the users, **the gestures for rotation and translation were inverted**, e.g., moving symmetrically two fingers on opposite sides of the cube triggered a rotation instead of the usual translation. Controls were only inverted for this task.
6. Positioning the road joining the first two pillars to the river bank with 2 translations and 1 rotation - repeated 3 times for different angles and starting position.
7. Positioning the road joining the four pillars with 1 translation, 3 rotations and scaling.

These different tasks enable us to observe how users get to learn how to use the CubTile for 3D objects manipulation tasks. Task number 5, with inverted control commands, enables us to observe mental workload while using voluntarily difficult and counter-intuitive interaction techniques. During the whole duration of the experiment, the participant brain and physiological signals were recorded.

## 3 Results

For each user, we first setup a workload level classifier based on the signals collected during the calibration session (N-back tasks). The next section describes the performances achieved for each participant and each signal type. Then, using the best workload classifier, we could estimate the workload level over time during the 3D docking tasks. This work was done offline, after the experiment.

**Accuracy of mental effort detection**

First, based on the data collected during the calibration session (N-back tasks), we could estimate how well low workload could be discriminated from high workload based on EEG, ECG and GSR. To do so, we used 2-fold cross-validation (CV) on the calibration data collected. In other words, we split the collected data into two parts of equal size, used one part to calibrate the classifier (CSP filters and sLDA) as described in section 2, and tested the resulting classifier on the data from the other part. We then did the opposite (training on the second part and testing on the first part), and averaged the obtained classification accuracies (percentage of signal time windows whose workload level was correctly identified). This CV was performed by using each signal type (i.e., EEG, ECG and GSR) either separately or in combination. Table 1 displays the obtained classification accuracies.

**Table 1.** Cross-validation classification accuracies (%) to discriminate workload levels from EEG, ECG and GSR on the calibration session data. A "*" indicates mean classification accuracies that are significantly better than chance ($p < 0.01$ according to [18])

| Participant | S1 | S2 | S3 | S4 | S5 | S6 | S7 | S8 | Mean |
|---|---|---|---|---|---|---|---|---|---|
| **EEG (Delta, Theta, Alpha)** | 74.0 | 76.2 | 76.5 | 80.2 | 84.9 | 81.9 | 81.7 | 75.4 | 78.9* |
| **EEG+EMG (Delta, Theta, Alpha, Beta, Gamma)** | 85.0 | 93.1 | 81.7 | 87.6 | 94.8 | 97.3 | 84.8 | 84.3 | 88.6* |
| **ECG** | 37.3 | 50.7 | 45.3 | 58.7 | 42.6 | 55.3 | 54.9 | 61.2 | 50.7 |
| **GSR** | 77.3 | 52.1 | 60.0 | 70.6 | 74.7 | 68.4 | 58.6 | 54.6 | 64.5 |
| **EEG+EMG+ECG+GSR** | 44.0 | 53.3 | 44.0 | 61.5 | 54.8 | 52.6 | 54.6 | 61.2 | 53.3 |

Classification results highlight that workload levels can be estimated in brain and physiological signals, even though the large inter-participant performance variability suggests that workload levels can be estimated more clearly for some users than for some others.

As can be first observed, it appears that EEG can discriminate workload levels with an accuracy higher than chance level, for all participants. In other words, the classification accuracies obtained are statistically significantly higher than 50% for a 2-class problem, i.e., more accurate than flipping a coin to estimate the workload level. Indeed, according to [18], for 160 trials per class, the chance level for $p < 0.01$ and a 2-class problem is an accuracy of 56.9%. Note that 180 trials per class were available with EEG in our experiment, meaning that the chance level is actually even slightly lower.

Regarding the GSR, it led to a better-than-chance classification accuracy only for some participants, but not for all. Indeed, we had 36 trials per class with GSR (due to the use of longer time windows as mentioned previously), which means a chance level of about 65% for $p < 0.01$ according to [18]. ECG signals could not lead to better-than-chance performances for any participant.

Overall, EEG seems to be the signal type the best able to discriminate workload levels reliably. Moreover, when EEG features include high frequency bands – i.e. when Delta, Theta and Alpha bands are combined with Beta and Gamma bands – and thus when EEG measures potentially contain EMG activity as well, the performances are the highest, close to 90% on average.

The poor performances of the system when all physiological signals are combined (EEG + EMG + ECG + GSR) may be explained by too important disparities in the features for the classifier to handle them correctly. On a side note, we also tested ECG and GSR on 2s time windows with adapted features – HR for the former, mean value and SCR for the latter. Despite the increased number of trials in training and testing phases, the results were very similar to those already described in Table 1. Altogether, the relatively poor performances obtained with ECG and GSR are likely due to the short time windows (2s or 10s long) used. Much better performances should be ex-pected with larger windows, e.g., with 30s-long or even 2min-long time windows [16], at the cost of a coarser temporal resolution.

Since we already obtained a classification accuracy close to 90% through the sole use of EEG recordings (which possibly include EMG activity as well), we did not push further our investigations about a multimodal (multiple signals) approach to mental effort estimation. Such method would necessitate longer time windows, strong synchronization between signals and extra classification steps, with little benefit to expect considering that a classifier based on GSR hardly reaches 65% of accuracy in our protocol.

We then calibrated the workload classifier on EEG signals from both low and high frequency bands (i.e., combining EEG and possibly EMG), and used it to analyze workload variations during the 3D manipulation tasks.

**Mental effort during 3D object manipulation**

While the participants were performing 3D docking tasks to build the 3D bridge, their brain signals were recorded. By using the workload level classifier obtained offline, such classifier being able to estimate whether the current 2-seconds long time window of signals corresponds to a low or high workload for the user, we could notably continuously estimate the workload levels during the tasks. This gave us unique insights into how much mental effort the participants were devoting to each task, and how this mental effort evolved over time.

Due to the large between-user variability in terms of workload level estimation accuracy, and since these estimations are not 100% accurate, we studied average workload levels to obtain a robust and reliable picture of the mental workload level associated with each task. To do so, we first normalized between -1 and +1 the output that was produced by the classifier for each participant during the virtual bridge construction. As such, a workload index close to +1 during the 3D object manipulation represents the highest mental workload a participant had to endure while performing the 3D docking tasks. It should come close to the 2-back condition of the calibration phase. In a similar manner, a workload index close to -1 denotes the lowest workload (similar to that of the 0-back condition).

Because there was no time constraint regarding task completion – users made as many attempts as needed to complete each one of them – we could not compare directly workload indexes over time. Some participants took more than 13 minutes to complete all the tasks while others finished in less than 5 minutes (mean: 7.7 min, SD: 2.9 min). This is why we averaged the workload index per task. Note that due to technical issues, for some participants the beginning and end of a couple of tasks were not accurately recorded or missing. If it was the case, the workload indexes for this task and participant were not included in the analysis to ensure unbiased results. Altogether, out of the 56 tasks (8 participants × 7 tasks per participant), 13 tasks were not included in the analysis to ensure clean results. More precisely, 1 task was missing for participant S3, 2 tasks were missing for participants S2 and S5, 3 tasks for participant S8, and 5 tasks for participant S4. No tasks were missing for the remaining partici-pants. We followed a rather conservative approach (i.e., we discarded a task in case of doubt), to ensure only clean and meaningful results are presented.

Figure 3 displays the workload levels averaged over all participants and over the duration of each docking task. This thus provides us with the average mental workload induced by each 3D object manipulation task.

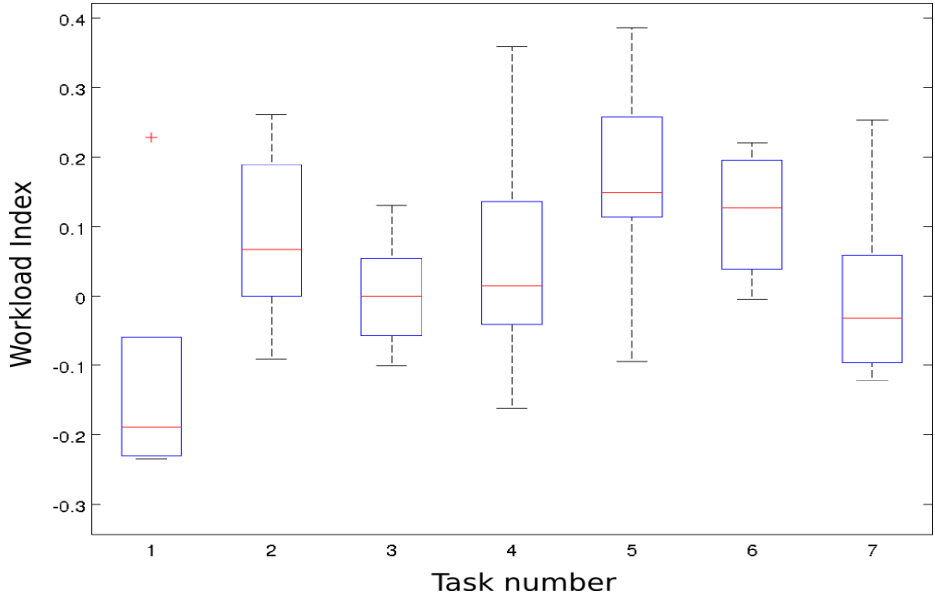

**Fig. 3.** Average workload levels (averaged over participants and task duration) measured for the different 3D docking tasks.

To ensure that the observed workload levels were really due to some information and structure in the data that are detected by the workload classifier, and not just due to chance or to some artefacts that are unrelated to workload levels, we performed a permutation analysis. In particular, we performed the exact same analysis described previously except that we used random classifiers instead of the real workload classifiers trained on the N-back task data. This aimed at estimating the type of workload level indexes we could obtain by chance on our data. To do so, for each participant, we shuffled the labels of the N-back task data, (i.e., the EEG signals were not labelled with the correct workload level anymore), and optimized the spatial filters and classifier described in Section 2 based on this shuffled training data. In order words, we built random classifiers that would not be able to detect workload levels. Then using these random classifiers for each participant, we computed the mean normalized workload level indexes for each 3D manipulation task, as described previously. We repeated this process (workload labels shuffling, then random classifier training, and testing of the classifier on the 3D manipulation task data) 1000 times, to obtain the distribution of the mean workload level indexes for each task that can be obtained by chance (see Figure 4). More precisely, we estimated the multivariate normal distribu-tion of the vectors of mean workload level per class (i.e., a vector with 7 elements, the $i^{th}$ element being the averaged workload level over participants for task i) obtained for each of the 1000 permutations. This multivariate distribution thus represents the mean workload levels per task that can obtain by chance. We finally compared the actual mean workload levels per task that we obtained using the real workload classifiers (i.e., those optimized on the unshuffled training data, whose output is displayed on Figure 3) to this chance multivariate distribution obtained with the random classifiers. This helped us estimate whether the obtained mean workload levels per task were due to chance or not. Results showed that the observed real workload levels are statistically significantly different from that obtained by the chance distribution with $p < 0.001$, i.e., they are not due to chance. This suggests that our workload classifier does find a workload level information during the 3D docking tasks that cannot be found by chance.

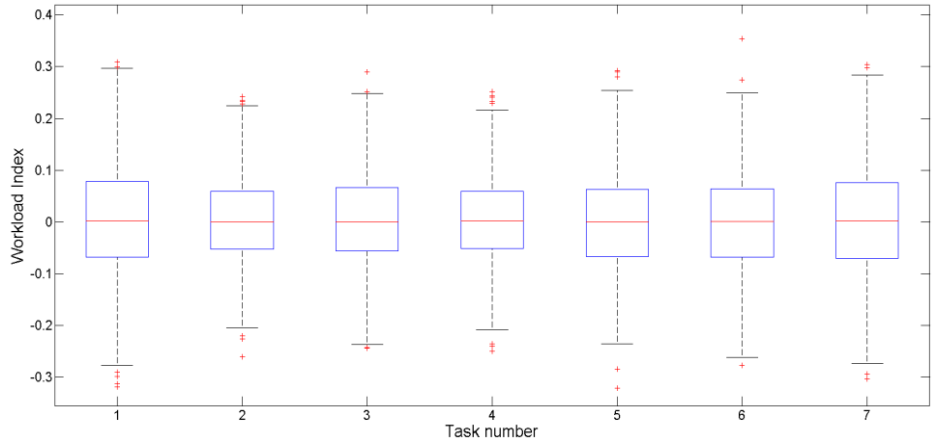

**Fig. 4.** Average workload levels obtained with a permutation test (see text for details), i.e., with random classifiers, for the different 3D docking tasks. The real workload levels we observed (i.e., those displayed in Figure 3) significantly differ from those random workload levels, i.e., they are not due to chance.

In order to sense whether or not the workload index fluctuated along tasks completion, we conducted a second analysis. Using the same normalized index, we compared the workload level between the first quarter and the last quarter of every task – average across tasks for each participant (Figure 5). A Wilcoxon Signed-rank test showed that there was no significant difference.

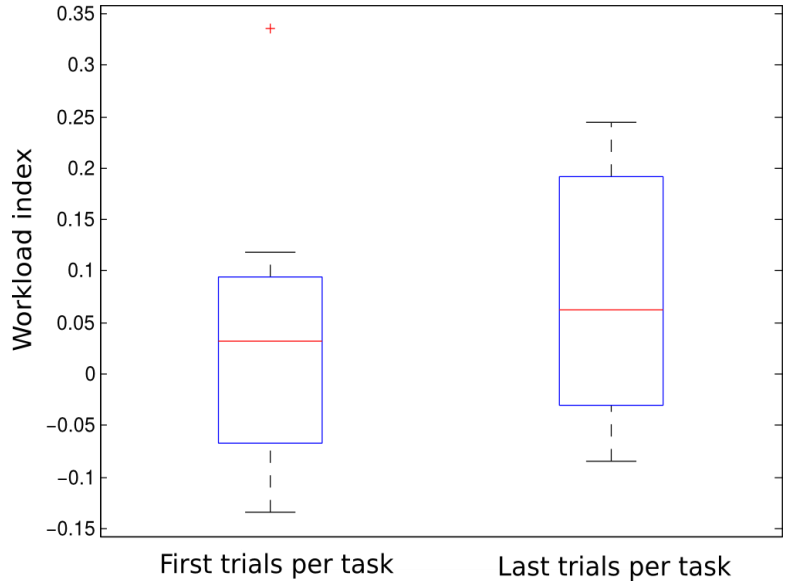

**Fig. 5.** Average per subject of the workload index during the first quarter of every task (left) compared to the last quarter (right).

## 4 Discussion

First, the fact that the observed workload levels during 3D manipulation tasks are not due to chance on the one hand and that our workload classifiers are calibrated on the N-back task, which is a widely used and validated workload induction protocol [21] on the other hand, strongly suggests that our approach may be used to observe how workload levels vary during 3D manipulation tasks. Indeed, our workload classifiers identified a specific EEG+EMG signature of workload levels thanks to the use of the N-back task, which then enabled us to estimate a non-random sequence of workload levels for each task. Naturally, if the variations of another mental state (or artefact) are highly correlated to that of the workload levels, and have a similar EEG+EMG signature as workload so that these variations are picked-up by our classifiers, then the observed variations of workload may be due to variations of another mental state. Therefore, the influence of a confounding mental state or artefacts cannot be completely ruled out without exploring how all possible mental states vary, which is of course impossible. However, the fact that our classifiers are specific to workload variations (since they are calibrated with the N-back task which specifically makes workload levels vary) and that the observed variations are not due to chance makes the influence of a such confounding factor rather unlikely. Based on this interpretation, we can now analyze how the workload level changes during the different 3D manipulation tasks and why.

The observed workload levels suggest that despite the novelty and the complexity of the interaction – handling at the same time rotation, translation and scaling of ele-ments in a 3D environment right from the beginning – the participants did not make an important mental effort to complete the first task. That could be due to the practicality of the CubTile, which may ease 3D interaction thanks to its additional degrees of freedom compared to a traditional input device such as a mouse.

When a constraint appeared concurrently with the second task – pillars were "falling" continuously from the sky and had to be positioned quickly before they touched the ground – the workload index increased substantially. This is consistent with the sudden pressure that was exerted on users. As one could expect, the mental workload lowered and settled in tasks 3 and 4, during which there was no more time pressure – but still more complex manipulations compared to task 1.

We purposely inverted the commands during the fifth task to disorientate participants. As a matter of fact, this is the moment when the workload index was the highest on average among all participants. Then, after this sudden surge of mental stress, once again the measured workload has been reduced in the two subsequent tasks. Interestingly enough, for task 6, in which the control commands were inverted back to normal, the workload indeed decreased as compared to that of task 5, but was still higher than for the other tasks. This probably reflects the fact that users had somehow integrated the counterintuitive manipulation technique and had to change again the gestures they used to manipulate the 3D object, thus being forced to forget what they had just learned in task 5 which resulted in a high workload. Since the new control scheme was the one they had already used during the previous tasks though, the workload was not as high as in task 5.

Overall, the mental workload that was measured with EEG and EMG along the course of the interaction matches the design of the tasks. Workload increased when a sensitive element of the interaction was deprived – e.g. time or commands – which can be explained by the need to overcome what participants have learned previously and re-learn how to handle the new environment. Afterward, when going back to the previous scheme, the workload goes back to a low level, as could be expected.

The absence of differences in the workload index between the beginning and the end of the tasks could be due to their durations. We expect to observe a learning effect when the CubTile – or any other input device – is operated during a prolonged period of time in steady conditions; i.e., the workload index would be lower in the end.

Overall, these results suggested that continuous mental workload monitoring was possible and could provide us with interesting insights about how cognitively easy-to-use a given 3D interaction technique can be. As compared to previous works, our results show that it is possible to monitor mental workload based on brain and physiological signals, even when the user is actively interacting (and not passively observing as in previous works), moving, and performing more complex and more cognitively demanding 3D manipulation tasks, in a visually rich 3D environment.

The approach we proposed here enabled us to perform continuous mental workload monitoring, but only with an offline analysis. Indeed, our algorithm required computing the covariance matrix of EEG signals recorded during the context of use (i.e., here during 3D object manipulation tasks), which would not have been possible if mental workload was to be estimated in real-time during these manipulation tasks. The covariance matrix was estimated on all the EEG data collected during the manip-ulation tasks, and thus could only be estimated once the tasks were completed. In the future, it would be interesting to design a continuous workload estimator that can also be used in real-time. To do so, our algorithm could be adapted in two ways: 1) the covariance matrix of the EEG signals recorded during 3D manipulation tasks could be estimated on the first task - or couple of tasks - only, to enable workload estimation in real-time on the subsequent tasks; 2) the differences between the calibration context and the use context are likely to be the same across different participants [27]. As such, the EEG signals directions that vary between contexts can be estimated on the data from some users, and used to estimate robustly the workload on the data from other users, hence without the need to estimate these variations for a new user, as done in [27] for the classification of EEG signals related to imagined hand movements. We will explore these options in the future, which would potentially open the door for robust continuous mental effort estimation during 3D interaction, in real-time.

## 5    Conclusion

In this paper, we have explored a new way to evaluate 3DUI in a more continuous, objective/exocentric and non-interrupting way. In particular we proposed to continuously monitor the mental effort exerted by users of a given 3DUI based on the measure of their brain signals (EEG). We first proposed a method to estimate such level of mental effort from EEG, EMG, ECG and GSR signals. We then used the resulting mental effort estimator to study mental workload during a pilot study involving 3D object manipulation tasks with a CubTile. Monitoring workload enabled us to continuously observe when and where the 3DUI and/or interaction technique was easy or difficult to use. In the future, it could potentially be used to also study how users learn to use the 3DUI, possibly in real-time. Overall, this suggested such approach can be a relevant tool to complement existing 3DUI evaluation tools.

Future works will consist in using the proposed workload estimator to assess other 3D interaction tasks such as navigation or application control. We will also explore other mental states that could be measured from brain and physiological signals, such as error recognition (to measure how intuitive a 3DUI can be) or emotions (to measure how pleasant and enjoyable a 3DUI can be). It would also be important and interesting to estimate whether and how wearing different sensors affects the way the user interacts with the 3DUI. Overall, we aim at designing a comprehensive evaluation framework based on brain and physiological signals that could be a new evaluation tool in the repertoire of UI designers.